# ■ High-Precision Value for the Quartic Anharmonic Oscillator Ground State

Michael Trott

Wolfram Research, Inc.
100 Trade Center Drive
Champaign, Illinois 61820-7237
mtrott@wolfram.com

We will describe how a new, quite simple, but highly effective algorithm, together with the asymptotically fast FFT-based high-precision number multiplication of *Mathematica* 4 can calculate the ground state of the $x^4$ anharmonic oscillator to the new record of more than 1000 digits.

## ■ Introduction

As it is well known, only a very limited number of one-dimensional potentials allow for an exact solution of the Schrödinger equation. This means that for many model potentials we must resort to numerical solution methods. For judging their accuracy, reliability, and speed, it is important to have high-precision values of certain nonexactly solvable potentials. The most investigated of such potentials is the quartic anharmonic oscillator (see [1] to [19]), described by

$$-\psi_k''(z) + z^4\, \psi_k(z) = \varepsilon_k\, \psi_k(z) \quad (1)$$

The eigenfunctions to the eigenvalues $\varepsilon_k$ decay exponentially for $z \to \pm\infty$.

## ■ The Hill determinant method

A classical method for solving Sturm-Liouville problems of type (1) is to calculate the eigenvalues of a truncated version of the corresponding Hill determinant. Using the harmonic oscillator basis $\phi_n(z)$ we write $\psi_0(z) = \sum_{k=0}^{\infty} \alpha_k\, \phi_k(z)$ where

$$-\phi_k''(z) + z^2\, \phi_k(z) = \epsilon_k\, \phi_k(z) \quad (2)$$

$$\phi_n(z) = \frac{1}{\sqrt{\sqrt{\pi}\, 2^n\, n!}}\, e^{-\frac{z^2}{2}}\, H_n(z) \quad (3)$$

Thus we form the matrix elements $h_{m,n} = \int_{-\infty}^{\infty} \phi_m(z)\, (-\phi_n''(z) + z^4\, \phi_n(z))\, dz$. For $n \geq m$ we obtain

$$h_{m,n} = \begin{cases} 0 & n-m > 4 \\ 2^{\frac{m-n}{2}-4}\, \sqrt{\frac{m!}{n!}}\, (32\, n\, \delta_{m,n-2}\, (n-1)^2 + 16\, (n-3)\, (n-2)\, n\, \delta_{m,n-4}\, (n-1) + \\ \quad 4\, (2\, n\, (3\, n+5) + 5)\, \delta_{m,n} + 8\, (n+1)\, \delta_{m,n+2} + \delta_{m,n+4}) & \text{else} \end{cases} \quad (5)$$

A rough estimation shows that we obtain about 0.2 digits per harmonic oscillator state. So, by taking into account the first 500 eigenstates and carrying out the calculation with about five thousand digits we obtain about 120 reliable digits for $\varepsilon_0$. (This calculation takes about 20 minutes on a year-2000 vintage workstation using *Mathematica* 4 [20].)

$$\varepsilon_0 = 1.06036209048418289964704601669266354551520872852897793321624524169594356304434444211268962991346717035105462443585825525580982763829\ldots$$

The Hill determinant approach allows the calculation of the eigenvectors in addition to the calculation of the eigenvalues. The following graphic visualizes the matrix of eigenvectors of $(h_{m,n})_{1 \leq n,m \leq 100}$. The graphic shows that the lowest eigenfunctions are quite similar to the harmonic oscillator eigenfunctions. Higher states are complicated mixtures of harmonic oscillator states. The overall "checkerboard"-like structure results from the fact that the contribution of the antisymmetric (symmetric) harmonic oscillator states to the symmetric (antisymmetric) anharmonic oscillator states is identically zero. The very high states are dominated by truncation effects and do not correctly mimic the anharmonic oscillator states.

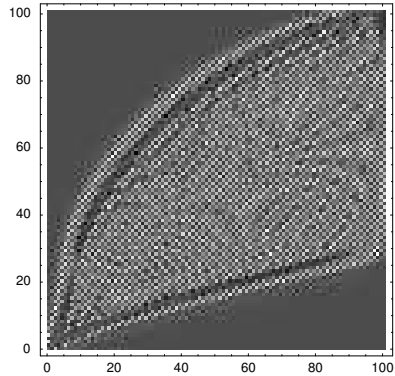

## ■ The new algorithm

To get a very high-precision approximation of

$$-\psi''(x) + z^4 \psi(x) = \lambda \psi(x) \quad (6)$$

we start with the series expansion

$$\psi(x) = y_n(x) = \sum_{k=0}^{n} a_k(\lambda) x^k \quad (7)$$

For the ground state we choose (ignoring normalization) $\psi(0) = 1$, $\psi'(0) = 0$. For "suitable chosen" $x^*$ we then find high-precision approximations for the zeros of $y_n(x^*)$ and $y'_n(x^*)$. These zeros then bound $\lambda_0$ from below and above.

Using the differential equation, one obtains the following recursion relation for $a_k(\lambda)$:

$$a_k(\lambda) = \frac{a_{m-6}(\lambda) - \lambda\, a_{m-2}(\lambda)}{m^2 - m} \quad (8)$$

For large $n$ ($n \to \infty$) we want the function $y_n(x)$ to vanish as $x \to \infty$. For a $\lambda$ smaller than the exact eigenvalue, the function $y_n(x^*)$ will not have a zero, but the function $y'_n(x^*)$ will have a zero for a certain $x^*$. For a $\lambda$ larger than the exact eigenvalue, the function $y_n(x^*)$ will have a zero, but the function $y'_n(x^*)$ will not have a zero for a certain $x^*$. This fact allows us to find a bounding interval for $\lambda_0$. The next two graphics show $y_{80}(x)$ and $y'_{80}(x)$ for 10 equidistant values for $\lambda$ from the interval [1.05, 1.08] to visualize this bounding process. (For more details, see [22].)

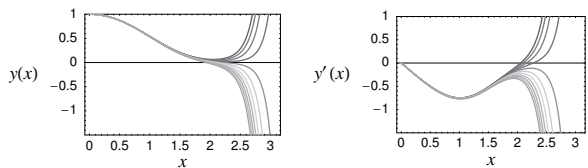

It is straightforward to implement the calculation of the bounding interval for $\lambda_0$ in *Mathematica* in a three-line program (see [21]). Using `FindRoot` we calculate high-precision values for the zeros of $y_n(\xi)$ and $y'_n(\xi)$.

```
λBounds[n_, ξ_, opts___] := Map[Function[f,
    λ /. FindRoot[f[n, λ, ξ] == 0, {λ, 106/100, 107/100}, opts]], {y, yPrime}];
```

The calculation of $y_n(\xi)$ and $y'_n(\xi)$ is also straightforward, based on recursive calculations of $a_k(\lambda)$.

```
y[n_, λ_Real, ξ_] := Module[{a6, a4, a2, ak, σ},
  {a6, a4, a2} = {1, -λ/2, λ^2/24}; σ = a6 + a4 ξ^2 + a2 ξ^4;
  Do[ak = a6 - λ a2 / (k (k - 1)); {a6, a4, a2} = {a4, a2, ak};
    σ = σ + ak ξ^k, {k, 6, n, 2}]; σ]

y[n_, λ_Real, ξ_] := Module[{a6, a4, a2, ak, σ},
  {a6, a4, a2} = {1, -λ/2, λ^2/24}; σ = a6 + 2 ξ a4 + 4 ξ^3 a2;
  Do[ak = a6 - λ a2 / (k (k - 1)); {a6, a4, a2} = {a4, a2, ak};
    σ = σ + k ak ξ^(k-1), {k, 6, n, 2}]; σ]
```

Now we calculate `λBounds[16000, 16, startingValues, WorkingPrecision→6000, AccuracyGoal→600, MaxIterations→100]` (where *startingValues* has been obtained from a call to `λBounds` recursively). In a few minutes, we get a 1184-digit approximation to the ground state energy of the quartic anharmonic oscillator.

$\varepsilon_0 =$
1.06036209048418289964704601669266345455152087285289779332162452416959435630443444211268962991346717037...
5105462443585825255808798082102931470131768363738249357892262460047081754469601416374884172822569037...
9357577908880617887902636015493956902751961489009429348735844094426948979012139714642909519233545337...
82834703350575761511202570398885237202402218411030865737310913989154536584103111679405833548600092...
74400696311267023886229714296996105921558322667137693550867361000083183002751792623357391390613618...
77649859696181499412792809272840707956106044072294680994913627572927387279136890279842472226171694...
48895475137043806840543918778772953234245874372543178323190603810687416044034374530146847278139186...
29404704310340135107160711035300892982327542766151898695056504716025275608952626219102568820096441...
28781564005270529293240507638265028259112477362538471854714402572285438485297450458570978840249066...
99570476844587709176202912437527325490711643344023029473069239819089568537453598844601600231329193...
05939586930491664428163394616332428700426146123774300995223420420859773569015356541685030894185134...
79573410658547971946759646679661346768858643795265451956056828671595833888474346701204242071491929...
048732…

Statistical analysis of the number does not show any regularity.

## ■ Summary

A power series-based approach to the high-precision calculation of the ground state of the anharmonic oscillator was presented. *Mathematica* code to carry out the calculation, as well as results, were given. The method can straightwardly be used to calculate tens of thousands of digits of the quartic anharmonic, as well as other anharmonic oscillators. Work concerning the application of the method to higher states is in progress.

All calculations and visualizations have been carried out in *Mathematica* 4.

## ■ Acknowledgments

The author would like to thank André Kuzniarek and Amy Young for making a prerelease version of the *Publicon* typesetting system (**www.publicon.com**). This work was supported by Wolfram Research, Inc.

## ■ References

[1] C. J. Tymczak, G. S. Japaridze, C. R. Handy, X.-Q. Wang. *quant-ph*/9707005.
[2] S. Graffi, V. Grecchi. *Phys. Rev.* D 8, 3487 (1973).
[3] A. V. Sergeev, D. Z. Goodson. *J. Phys.* A 31, 4301 (1998).
[4] J. Suzuki. *J. Phys.* A 32, L183 (1999).
[5] L. V. Chebotarev. *Ann. Phys.* 273, 114 (1999).
[6] P. E. Shanley. *Ann. Phys.* 186, 292, 325 (1988).
[7] T. Hatsuda, T. Kunihiro, T. Tanaka. *Phys. Rev. Lett.* 78, 3229 (1997).
[8] E. J. Weniger. *Phys. Rev. Lett.* 77, 2859 (1996).
[9] K. C. Ho, Y. T. Liu, C. F. Lo, K. L. Liu, W. M. Kwok, M. L. Shiu. *Phys. Rev.* A 53, 1280 (1996).
[10] J. Čizek, E. J. Weniger, P. Bracken, V. Špirko. *Phys. Rev.* E 53, 2925 (1996).
[11] K. Bay, W. Lay. *J. Math. Phys.* 38, 2127 (1997).
[12] A. Voros. *J. Phys.* A 32, 5993 (1999).
[13] P. B. Khan, Y. Zarmi. *J. Math. Phys.* 40, 4658 (1999).
[14] F. Antonsen. *Phys. Rev.* A 60, 812 (1999).
[15] L. Skála, J. Čizek, V. Kapsa, E. J. Weniger. *Phys. Rev.* A 56, 4471 (1997).
[16] H. Meißner, E. O. Steinborn. *Phys. Rev.* A 56, 1189 (1997).
[17] E. Delabaere, F. Pham. *Ann. Phys.* 261, 180 (1997).
[18] F. M. Fernández, R. Guardiola. *J. Phys.* A 26, 7169 (1993).
[19] F. Vinette, J. Čizek. *J. Math. Phys.* 32, 3392 (1991).
[20] S. Wolfram. *The Mathematica Book, 4th Edition*, Cambridge University Press and Wolfram Media, 1999.
[21] M. Trott. *The Mathematica GuideBook: Programming in Mathematica*, Springer-Verlag, New York, 2001.
[22] M. Trott. *The Mathematica GuideBook: Mathematics in Mathematica*, Springer-Verlag, New York, 2001.